\documentclass[aps,pra,showpacs,notitlepage]{revtex4-1}
\usepackage{graphicx}

\begin{document}

\title{Quantum states as complex probabilities:\\
The physics behind direct observations of\\ photon wavefunctions in weak measurements}

\author{Holger F. Hofmann}
\email{hofmann@hiroshima-u.ac.jp}
\affiliation{
Graduate School of Advanced Sciences of Matter, Hiroshima University,
Kagamiyama 1-3-1, Higashi Hiroshima 739-8530, Japan}
\affiliation{JST, CREST, Sanbancho 5, Chiyoda-ku, Tokyo 102-0075, Japan
}

\begin{abstract}
Weak measurements of photon position can be used to obtain direct experimental evidence of the wavefunction of a photon between generation and ultimate detection. Significantly, these measurement results can also be understood as complex valued conditional probabilities of intermediate photon positions. It is therefore possible to interpret the quantum state as a complex valued probability distribution from which measurement probabilities can be derived according to Bayesian rules. The conventional measurement probabilities derived from squares of the wavefunction then describes the effects of measurement back-action, which originate from a non-classical relation between dynamics and statistics that is characteristic of quantum mechanics. It is pointed out that this relation can be used to derive the complete Hilbert space formalism directly from complex probabilities, without the axiomatic introduction of quantum states or operators.
\end{abstract}

\maketitle

\section{Introduction}

Quantum mechanics describes the dynamics of a particle in space and time in terms of the evolution of a wavefunction $\psi(x)$. A precise measurement of $x$ will find the particle at a specific position given by a single value of $x$, but quantum coherence between different positions $x$ is necessary to define the momentum of the particle, and hence the direction in which the particle will move after passing the intermediate position $x$ on its way from an initial preparation to its final detection. 

All interpretational problems of quantum mechanics arise from the fact that we cannot obtain information on the precise position $x$ of a particle that moves from an initial state $\psi$ to a final measurement of momentum $p$. Specifically, textbook quantum mechanics fails to provide a physical justification of the expression $\psi(x)$, since the attachment of the amplitude $\psi(x)$ to a specific position $x$ is not explained in terms of a corresponding measurement of $x$. It is therefore significant that recent advances in quantum optics indicate that it is in fact possible to obtain direct experimental evidence of $\psi(x)$ for particles propagating from $\psi$ to $p$: as shown by Lundeen and coworkers in 2011, the wavefunction of a photon can be obtained by performing a weak measurement of the projector $\mid x \rangle\langle x \mid$ in the focal plane, followed by a post-selection of the $p=0$ result associated with a final detection of the photon at the center of the Fourier plane \cite{Lun11}. This measurement shows that the wavefunction $\psi(x)$ is proportional to the weak value of the probability of finding the photon at $x$. It is therefore possible to interpret the wavefunction as a complex conditional probability $p(x|\psi,p=0)$ that relates the initial and final conditions $(\psi,p=0)$ to the intermediate position $x$. In the following, I will argue that this interpretation of the quantum state provides a complete and consistent explanation of quantum physics that can resolve the fundamental questions concerning the status of microscopic realities in quantum systems \cite{Pus12,Hof11a,Hof13}. 

\section{Weak measurement of photon wavefunctions}

Weak measurements can determine the value of a physical property $\hat{A}$ between preparation and measurement without disturbing the dynamics of the quantum system. By averaging over a large number of measurements obtained with the same intial and final conditions $\mid \psi \rangle$ and $\mid f \rangle$, one obtains the weak value \cite{Aha88}, 
\begin{equation}
\langle \hat{A} \rangle_{\mathrm{weak}} (\psi,f) = \frac{\langle f \mid \hat{A} \mid \psi \rangle}{\langle f \mid \psi \rangle}.
\end{equation}
It might be worth noting that the mathematical form of weak values was discovered as early as 1945 by Dirac, who derived it as the most simple way to express an observable $\hat{A}$ as a function of pairs of eigenvalues from two non-commuting physical properties, in close analogy to the classical expression of properties as a function of the position and momentum coordinates in phase space \cite{Dir45}. Dirac also noted that the same formalism can be applied to represent the density operator of a quantum state as a complex joint probability of non-commuting properties. Thus, weak measurements can be understood as an experimental method of determining the complex probabilities predicted by Dirac more than 30 years before the proposal of Aharonov, Albert and Vaidman, and more than 60 years before the first direct measurement of the quantum state as a complex probability by Lundeen and coworkers \cite{Joh07,Hof12,Lun12}.

In the context of wavefunction measurements, the weak value of the probability that a photon prepared in $\psi$ and detected at $p=0$ in the Fourier plance passed through $x$ in the focal plane is given by
\begin{equation}
p(x|\psi,p=0) = \frac{\langle p=0 \mid x \rangle \langle x \mid \psi \rangle}{\langle p=0 \mid \psi \rangle}.
\end{equation} 
As observed by Lundeen and coworkers, the $x$-dependence of this complex conditional probability is identical to the $x$-dependence of the wavefunction $\psi(x)=\langle x \mid \psi \rangle$ \cite{Lun11}. In fact, the only difference is in the normalization: since the total conditional probability should be one,
\begin{equation}
p(x|\psi, p=0) = \frac{\psi(x)}{\int \psi(x^\prime) dx^\prime}.
\end{equation}
It may therefore be possible to interpret the wavefunction more directly as a complex conditional probability, in close analogy to classical determinism \cite{Hof12}. Specifically, the classical notion that the path of a particle from $\psi$ to $p=0$ should uniquely determine the intermediate positions $x$ is now replaced with the notion that measurements of $x$ in the focal plane are fundamentally related to alternative measurements of $p$ in the Fourier plane by complex conditional probabilities of the form $p(x|\psi,p)$ \cite{Hof13}. 

\section{Quantum ergodicity}

To understand the meaning of complex conditional probabilities better, it is useful to consider the only accessible physics described by the wavefunction $\psi(x)$, which is the measurement probability of $x$ conditioned by $\psi$, $p(x|\psi)=|\psi(x)|^2$. When expressed in terms of complex conditional probabilities, this relation is modified to 
\begin{equation}
\label{eq:qergod}
p(p|x)p(x|\psi) = p(p|\psi) |p(x|\psi,p)|^2.
\end{equation}
Here, the probabilities on the left hand side represent the statistics of a sequential measurement of $x$ and $p$, where the back-action of the $x$-measurement completely randomizes the dynamics along $x$, erasing any effects of the property $\psi$ on the final outcome $p$. The right hand side shows that the relation of this sequential measurement to the direct measurement probability of $p$ in $\psi$ can be represented by the absolute values of the complex conditional probabilities. 

In general, the decoherence effects of measurement interactions can be explained by a randomization of the dynamics conserving the target observable. Thus, the measurement probabilities $p(b|a)$ obtained in projective measurements of $b$ represent the ergodic probabilities of $b$ obtained from averages along the dynamical trajectory defined by the initial property $a$. Eq.(\ref{eq:qergod}) shows how the fundamental relation between the three properties $\psi$, $x$, and $p$ observed in weak measurements with negligible measurement back-action relates to the ergodic probabilities observed in strong measurement with maximal measurement back-action. Significantly, Eq.(\ref{eq:qergod}) is independent of the technical circumstances of the measurements and refers only to the system itself. It can therefore be regarded as a universal law that defines the fundamental relations between the different properties of physical systems in the ultimate microscopic limit. Since this law essentially relates the deterministic relations between the physical properties expressed by complex conditional probabilities to the effects of dynamical averaging, I propose to refer to this law as the law of quantum ergodicity \cite{Hof13}.

\section{Determinism of complex probabilities}

Before explaining the connection between the complex phases of conditional probabilities and the dynamics associated with measurement interactions in more detail, it may be necessary to explain why complex conditional probabilities may actually be used to express fundamental relations between physical properties. Specifically, it is important to understand that the relations expressed by complex probabilities can be deterministic, even though there is no joint reality shared by the three different properties in this relation. 

Firstly, it should be quite clear that complex probabilities do not represent relative frequencies. Since they refer to alternative measurements that cannot be performed jointly, there exists no measurement outcome $(a,b)$ that would assign simultaneous reality to both $a$ and $b$ in the complex joint probability $\rho(a,b)$ describing a (possible mixed) quantum state $\rho$. However, the actual measurement probability $p(m)$ for a completely different property $m$ can be obtained from $\rho(a,b)$ by using Bayesian rules \cite{Hof12},
\begin{equation}
p(m)=\sum_{a,b} p(m|a,b) \rho(a,b).
\end{equation}
In this expression of the actual measurement probability $p(m)$, the complex conditional probability $p(m|a,b)$ is the universal relation between $m$, $a$ and $b$, valid for all possible states $\rho$ and independent of context or circumstance. Complex conditional probabilities are thus more fundamental than the related statistical expressions for quantum states. 

A quantitative test of the fundamental nature of the relation $p(m|a,b)$ can be obtained by observing that $(a,b)$ and $(m,b)$ should both provide an equally complete description of all quantum states. A transformation from $(a,b)$ to $(m,b)$ and back to $(a,b)$ should therefore return the same value of $a$, based on the relations $p(m|a,b)$ and $p(a|m,b)$ that describe the transformations. The mathematical expression for this condition reads 
\begin{equation}
\label{eq:determinism}
\sum_m p(a^\prime |m,b) p(m|a,b) = \delta_{a,a^\prime}.
\end{equation}
Complex conditional probabilities are deterministic if, and only if, they fulfill this requirement. 

It is now possible to re-formulate quantum ergodicity in terms of the contributions of each measurement result $m$ to the deterministic relation for $a=a^\prime$. Specifically,
\begin{equation}
\label{eq:qergo2}
p(a|m,b) p(m|a,b) = p(m|a).
\end{equation}
In this expression of the law of quantum ergodicity, the fact that the right hand side is real requires that the complex phases of $p(m|a,b)$ and of $p(a|m,b)$ are exactly opposite, highlighting the fact that the sign of the phase depends on the cyclic ordering of $a$, $m$, and $b$. Additionally, it is clear that the effects of $b$ on the complex probabilities on the left hand side cancel each other along with the complex phases. As discussed in \cite{Hof13}, this cancellation can be used to explain the origin of the Hilbert space inner product in terms of the more fundamental relations between three physical properties of the quantum system given by complex conditional probabilities.

\section{Origin of the Hilbert space formalism}

Eqs.(\ref{eq:determinism}) and (\ref{eq:qergo2}) show that the different outcomes $a$ can be described by normalized orthogonal vectors $\mid a \rangle$ with components given by the inner product $\langle m \mid a \rangle$ with the basis-outcomes $m$, where the reference $b$ is only required for the definition of the complex phases of the inner products $\langle m \mid a \rangle$. The Hilbert space algebra can thus be obtained from the experimentally accessible complex conditional probabilities by \cite{Hof13}
\begin{equation}
\langle m \mid a \rangle = \sqrt{\frac{p(b|a)}{p(b|m)}} p(m|a,b).
\end{equation}
Importantly, the phase reference $b$ eliminates all arbitrary phases and explains the physics behind gauge transformations in quantum mechanics. The omission of this third property in the original formulation of Hilbert space theory was an inadvertent consequence of the historical derivation, which was not properly rooted in empirical physics and relied on mathematical speculation instead. The empirical derivation of quantum mechanics from weak measurement statistics solves this problem and shows that all phases in quantum mechanics have a proper physical meaning. 

\section{Correlations of dynamics and statistics}

As indicated in the discussion above, complex conditional probabilities provide a more detailed explanation of the physics described by quantum theory. In this explanation, the difference between quantum mechanics and a non-quantum description of the ``classical'' world originates from the replacement of deterministic relations between physical properties with complex conditional probabilities. To understand the essence of quantum mechanics, it is therefore necessary to understand the role of the complex phases of conditional probabilities in the classical limit. 

In weak measurements, the imaginary parts of weak values are actually obtained from the response of the post-selection probabilities on weak unitaries generated by the observable of the weak measurement \cite{Hof11,Dre12}. Weak measurements thus blur the distinction between the value of a physical property and the dynamical change generated by the same property. The mathematical form of weak values suggests that this elimination of the distinction between static realities and dynamic transformations may actually be the essence of quantum mechanics. 

As explained in \cite{Hof11}, the phases of complex probabilities $p(m|a,b)$ express the action of an optimal transformation from $a$ to $b$ along constant $m$, where the fundamental constant $\hbar$ is the ratio between the action and the phase. Quantum mechanics thus indicates that the concept of an instantaneous and static reality described by a spatial geometry at time $t$ is an approximation that breaks down in the limit of small actions, since reality can only be experienced in the form of interactions that also change the system. Thus, the separate realities of different physical properties are related to each other by complex conditional probabilities, where the complex phases represent a description of the dynamics by which one reality can be transformed into another without disturbing the third. 

At the macroscopic level, we do not usually resolve the small imaginary correlations that describe the transformations between the observation of different physical properties. However, there exists one well known relation that directly identifies the macroscopic effects of imaginary probabilities. This relation is the quantum mechanical description of the dynamics of expectation values,
\begin{equation}
\frac{d}{dt} \langle \hat{A} \rangle = - \frac{i}{\hbar} \langle [\hat{A},\hat{H}] \rangle
= \frac{2}{\hbar} \; \mbox{Im}\left(\langle \hat{A}\hat{H} \rangle \right).
\end{equation}
Thus, quantum mechanics identifies the time evolution of any observable $\hat{A}$ with the imaginary part of the correlation between that observable and the energy, multiplied with a constant factor of $2/\hbar$. Any change in $A$ requires imaginary joint probabilities of $A$ and the energy $E$, and the rate of change observed can immediately be interpreted as the magnitude of the imaginary correlation. In this sense, complex probabilities are actually a part of everyday experience. For example, the imaginary correlation between position and energy for any object moving at a velocity of $\mbox{v}$ is $\hbar \mbox{v}/2$, no matter how large the object is. Perhaps it would be worthwhile to reflect on this simple relation in order to understand just how unrealistic it is to extrapolate the classical laws of motion to their microscopic limit of mathematical lines in space and time.  

\section{Conclusions}

Quantum mechanics has come a long way. In the recent decades, new possibilities of measurement and control have emerged, revealing the full potential of non-classical relations between physical properties that cannot be measured at the same time. Even while the philosophical discussions about the proper interpretation of the quantum formalism has gained new importance, recent experimental demonstrations of weak measurements have shown that there are also empirical methods that can reveal more about the physics described by quantum states. In particular, the direct measurement of the wavefunction by Lundeen and coworkers has revealed a fundamental and previously unknown relation between the Hilbert space formalism and complex valued Bayesian probabilities \cite{Lun11}. 

The explanations above are intended to show how this new experimental evidence can be used to formulate a more consistent explanation of quantum mechanics, based on the relations between alternative measurements revealed by the complex conditional probabilities obtained in weak measurements. Significantly, the complex phases of the conditional probabilities express the dynamical relation between alternative measurements in terms of the action of transformation. At the most fundamental level, quantum mechanics thus describes the dependence of reality on the dynamics by which it emerges in the course of a measurement interaction. The misconceptions that have resulted in the interpretational confusion in quantum mechanics originate from the wrong assumption that reality should be static and independent of dynamical effects. However, empirical reality always requires an interaction by which the reality of the object can be exprienced as a force acting on its environment. Quantum mechanics incorporates the structure of these forces in the complex conditional probabilities, as described by the law of quantum ergodicity.

The law of quantum ergodicity is the foundation of the conventional Hilbert space formalism \cite{Hof13}. It is therefore possible to base all quantum mechanical descriptions of physical systems on the law of quantum ergodicity, which provides a universal prescription of how all laws of physics can be formulated in their appropriate quantum form. Eventually, the law of quantum ergodicity and the use of complex conditional probability for the expression of deterministic relation between different physical properties should thus result in a much deeper understanding of quantum mechanics as a fundamental theory of physics.

\end{document}